\documentclass[pra,preprint]{revtex4}

\def\>{\rangle}
\def\<{\langle}

\usepackage{graphicx,color}
\usepackage{amssymb}
\usepackage{makeidx}
\begin{document}

\title{Influence of configuration and acoustic phonons on three-qubit controlled-phase gate}
\author{Song Yang}
\author{Ming Gong}
\author{Xubo Zou\footnote{xbz@ustc.edu.cn}}
\author{Chuanfeng Li}
\author{Cheng Zhao}
\author{Guangcan Guo}
\affiliation{Key Laboratory of Quantum Information, University of
Science and Technology of China, Hefei, 230026, People's Republic of
China}

\date{\today}

\begin{abstract}
We discuss the configuration and acoustic phonon effect on the
conditional phase gate using self assembled quantum dots. As an
example, we discuss the simplest three dots conditional phase gate,
and we found that the fidelity of conditional phase gate depends
strongly on the configuration and temperature. Numerical simulation
shows that line-configuration resonant with lower single exciton
energy level performs better during the gate operation.
\end{abstract}

\maketitle

The implementation of a sequence of quantum gate operations and the
ability to initialize the qubits are prerequisites for quantum
information processing (QIP).  So far, the two-qubit controlled-NOT
(CNOT) and controlled PHASE (CPHASE) gates have been demonstrated
experimentally in cavity\cite{turchette}, ion traps\cite{monroe},
NMR\cite{jones}, quantum dot (QD)\cite{ljsham} and superconductor
system\cite{yamamoto}, and the three-qubit CPHASE gate in NMR system
has also been reported recently\cite{vanderwal}. However, among all
the realization schemes, the electron spin in QD attracts special
interest due to the integrated advantages: (i), the spin in QD can
be manipulated ultrafast by a circularly polarized pulse ($\sim$
ps), which may be more readily realized with current technology.
(ii), the QDs can be scaled up to large network benefitted from the
state-of-the-art electronics. (iii),the spin decoherence lifetime is
very long\cite{cor02,imamo}, so numerous operations can be performed
before the coherence is totally lost during the interaction between
QDs and the phonon bath.

Recently, a scheme to realize CPHASE gate using QD molecular has
been proposed by Gauger{\it  et al}\cite{erik08, erik08a}, who shows
that the phonon bath has huge influence on the CPHASE. Here, we
extend this model to three QDs system, which is of great importance
in QIP. In this letter, we focus on the configuration effect on the
CPHASE operation.

\begin{figure}
\includegraphics[width=3.2in]{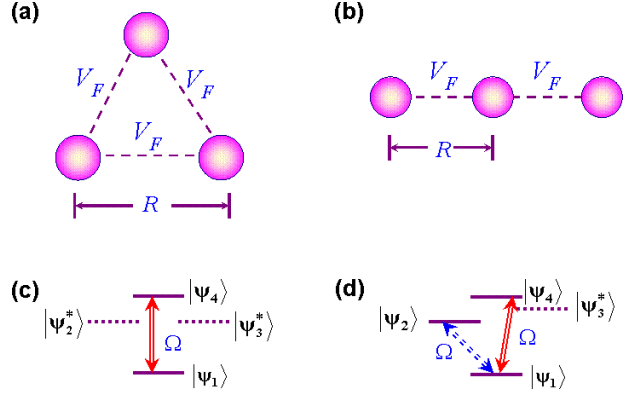}\\
\caption{(color online). Possible configuration of three quantum
dots, ring for (a) and line for (b). We assume that the three
quantum dots have identical exciton energy and have the same exciton
transfer rate. Due to F\"{o}rster interaction between the coupled
QDs, energy shift on single-exciton states in subspace $\mathcal{H}$
occurs. We can choose an appropriate single-exciton state as
auxiliary state to achieve CPHASE gate. The energy level diagrams of
ring and line configuration are shown in (c) and (d) respectively.
The arrows denote the possible resonant transitions to implement
gate operation: only $|\Psi_1\rangle\leftrightarrow|\Psi_4\rangle$
is allowed in ring structure, both
$|\Psi_1\rangle\leftrightarrow|\Psi_2\rangle$ and
$|\Psi_1\rangle\leftrightarrow|\Psi_4\rangle$ are possible in line
structure. } \label{fig:scheme}
\end{figure}

The system we studied is composed of three lateral coupled  QDs, as
schematically shown in Fig. \ref{fig:scheme}, (a) is ring
configuration and (b) is the line configuration. We use electron
spins to realize the CPHASE because  the spin has sufficiently long
decoherence time \cite{borri01}.  In self-assembled QDs, the lattice
mismatch between the QD and the substrate break the degenerate of
heavy hole (HH) and light hole (LH) by about 200 meV, hence the HH
can be well described by $\pm {3\over 2}$. The tunneling effect
decreases quickly as the distance between dots increases, and can be
neglected. The hamiltonian to describe this system hence read as

\begin{equation}
H_0=\omega_{a}\sum_{i}n_i+V_F\sum_{\langle i,j\rangle
}c_i^{\dagger}c_j+V_{xx}\sum_{\langle i, j\rangle}n_in_j,
\label{eq:Ham}
\end{equation}
where $c_i$ and $c_i^+$ is the destruction and creation operator of
exciton at the i$^{\text{th}}$ QD. $V_F$ is the F\"{o}rster energy
transfer rate between coupled QDs. It is dominated by the
dipole-dipole interactions, and has recently been demonstrated by
Kim {\it et al} in semiconductor QDs\cite{kim08}. The third term
$V_{xx}$ is the Coulomb interaction between  the coupled QDs, and
may contribute to the system when each QD contains one exciton.

We now impose a time-dependent and $\sigma^+$ polarized laser pulse
to realize the three QDs CPHASE gate. The Hamiltonian can be written
as
\begin{equation}
H_i = \Omega\cos(\omega_l t) \sum_i (c^{\dag}_{i}+  c_i)
\end{equation}
where $\omega_l$ is the frequency of the external field, and
$\Omega$ is the corresponding coupling constant. Because the
external field is $\sigma^+$ polarized, so only
$|X^+_{\Uparrow}\rangle=|\Uparrow\downarrow\rangle$ is possible to
be created when the qubit is initialized to
$|\uparrow\rangle$\cite{rikitake06}. If the qubit is initialized at
$|\downarrow\rangle$, no exciton will be created in the QDs because
of the Pauli blocking effect. With the above restriction, the
hamiltonian can be decoupled into four subspaces sorted by the
number of qubit in the basis state $|\uparrow\rangle$. Additionally,
with the help of F\"{o}rster interaction $V_{F}$ and Coulomb
interaction $V_{xx}$, the energy shifts lifted in each subspace are
different. We are primed to pay attention to the subspace
$\mathcal{H}$ with its electron state
$|\uparrow\uparrow\uparrow\rangle$, whose level schemes as shown in
Fig. \ref{fig:scheme} (c) (ring configuration) and Fig.
\ref{fig:scheme} (d) (line configuration).

{\it One-step CPHASE gate.}--Most three-qubit phase gate requires a
lot of two-qubit gates and one-qubit rotations\cite{diao02}, or
performing many steps by addressing individual qubit with laser
pulses\cite{jaksch00}. In this work, the CPHASE gate can be realized
in only one step. First, we will study the performance of the CPHASE
gate in the ideal case without any dephasing mechanism. We derive
Eq.(\ref{eq:Ham}) into the subspace $\mathcal{H}$ with its eight
eigenvectors. We drive the system with a $\sigma^{+}$ laser field
tuned on resonance with the transition
$|\uparrow\uparrow\uparrow\rangle\leftrightarrow|\Psi_s\rangle$, and
$|\Psi_s\rangle$ is an auxiliary state. Assuming other off-resonant
transitions are so far off the resonance that no transition is
induced through it, we can get an effective Hamiltonian
\begin{equation}
H_{\text{eff}} = {\alpha\Omega \over 2} |\Psi_s\rangle \langle
\uparrow\uparrow\uparrow| +h.c \label{eq:effective1},
\end{equation}
in which $\alpha$ is character parameter which is determined by its
configuration. Transforming back to the lab frame, the time
evolution of the initial state $|\Psi_ 1\rangle$ may be written as
\begin{eqnarray}
|\uparrow\uparrow\uparrow\rangle\rightarrow\cos[\theta(t)]|\uparrow\uparrow\uparrow\rangle+e^{-i\omega_{l}t}\sin[\theta(t)]|\Psi_
s\rangle
\end{eqnarray}
where $\theta(t) = \frac{\alpha}{2}\int dt \Omega$. In this way,
through the resonant interaction Eq.(\ref{eq:effective1}), one can
choose the interaction time $t$ with $\theta(t)=\pi$, the state
$|\uparrow\uparrow\uparrow\rangle$ will acquire an $e^{i\pi}$ phase
factor, i.e.,
\begin{eqnarray}
|\uparrow\uparrow\uparrow\rangle\rightarrow-|\uparrow\uparrow\uparrow\rangle\
\end{eqnarray}
The polarized laser field can not induce transitions through initial
states to exciton states in any other subspaces. The state
$|\downarrow\downarrow\downarrow\rangle$ can not evolve under
$\sigma^{+}$ polarized light for the Pauli blocking principle. And
electronic states owing single $|\uparrow\rangle$ and that owing
double $|\uparrow\rangle$ are off-resonant with their single-exciton
states via the large detuning $\Delta$, which is tuned large enough
such that $\Delta\gg\Omega$. Therefore, the classical light field
combing the coupling between the dots can create a $\pi$ phase shift
on the state $|\downarrow\downarrow\downarrow\rangle$ and leave the
other seven electronic state unevolved, implementing the three-qubit
CPHASE gate.

{\it Dephasing due to the phonon bath.}--In the actually phase gate,
the coupling with the phonon bath can lead to dephasing, which
 degrade the quality of the phase gate. We focus on the low
temperature regime, hence only the acoustic phonons will contribute
to the dephasing precess\cite{ben07,krum02}.   The exciton phonon
interaction can be described by

\begin{equation}
H_{ep}=\sum\limits_{j=1}^{3}\sum\limits_{\mathbf{q}}^{
}g_{q,j}c_j^{\dag}c_j(a_{\mathbf{q}}+a_{\mathbf{q}}^{\dag})
\end{equation}
with the effective excitonic coupling strength
\begin{equation}
g_{\mathbf {q},j}=\sum\limits_{\mathbf{q}}^{ }e^{{i\mathbf{q}\cdot
\mathbf{d_{j}}}}[M_{q,j}^{e}\rho_{e}(\mathbf{q}) -M_{q,j}^{h}
\rho_{h}(\mathbf {q})],
\end{equation}
 where  $M_{q,j}^{e/h}=\sum\limits_{\mathbf{q}}\sqrt{\hbar|q|/2\mu V
c_{s}}D_{e/h}$, and the state density in $\mathbf{k}$ space
$\rho_{e/h}(\mathbf {q})=\int
d^{3}r|\phi_{e/h}|^2e^{i\mathbf{q}\cdot\mathbf{r}}$. $a_{\mathbf
q}$($a_{\mathbf{q}}^{\dag}$ ) are the annihilation(creation)
operators for phonons with wave vector ${\bf q}$, $\mu$ denotes the
mass density, $V$ is the normalization volume, and $D_{e(h)}$ is the
deformation potential coupling constant of elector (hole). The wave
function we choose is $\phi_{e/h} \sim\exp(-r^2/l_{e/h}^2)$with the
electron(hole) ground-state localization length. Because the wave
function has spherical symmetry, only the longitudinal acoustic
phonon (LA) will contribute to the decoherence process. Under the
Markovian approximation, the master equation of the  density for the
whole system can be read as in a Lindblad form\cite{erik08}

\begin{eqnarray}
\dot{\rho}&=&-i[H_s,\rho]+\mathcal{R}[\rho]+\sum_{i}J(\omega_{i})([N(\omega_{i})+1]D[L_{i}]\rho\nonumber\\
&+& N(\omega_{i})D[L^{\dag}_{i}]\rho). \label{eq:ME}
\label{eq:master}
\end{eqnarray}
with $H_s = H_0 + H_i$, and $\emph{R}[\rho]=
\sum\limits_{i=1}^{3}\Gamma[\sigma_{i-}\rho\sigma_{i+}-\frac{1}{2}(\sigma_{i+}\sigma_{i-}\rho+\rho\sigma_{i+}\sigma_{i-})]$
 depicting the photon emission effect and $\sigma_{i-}=|\uparrow\rangle_{i}\langle
X^+_{\Uparrow}|$ (or $\sigma_{i+}=|
X^+_{\Uparrow}\rangle_{i}\langle\uparrow|$) denoting the lowing (or
rasing) operators of $i^{th}$ dot. $\Gamma$ is the spontaneous
radiative decay rate, which is inversely proportional to the
lifetime of exciton. $D[L]\rho=L\rho L^{\dag}-{1\over
2}(L^{\dag}L\rho+\rho L^{\dag}L)$ is the decay operator of phonon
effect, and $N(\omega)=[\exp(\omega/ k_{B}T)-1]^{-1}$ is the thermal
occupation of the phonon modes. $J(\omega_{i})$ is the phonon
spectral density, which is determined by the configuration. Here we
mainly focus on the widely studied InAs/GaAs QDs, and we choose the
parameter $V_F = 0.85$ meV, Exciton energy $\omega_a = 1.1$ eV,
bi-exciton binding energy $V_{xx} = 5$ meV, radiative decay rate
$\Gamma =1.6$ $\mu$ eV, electron and hole ground state localization
length $l_e = 2.16$ nm, $l_h = 1.44$ nm, mass density
$\mu=5.3$k/cm$^3$ and the sound velocity $c_s = 4.8\times 10^5$
cm/s.

\begin{figure}
\centering
\includegraphics[width=3.2in]{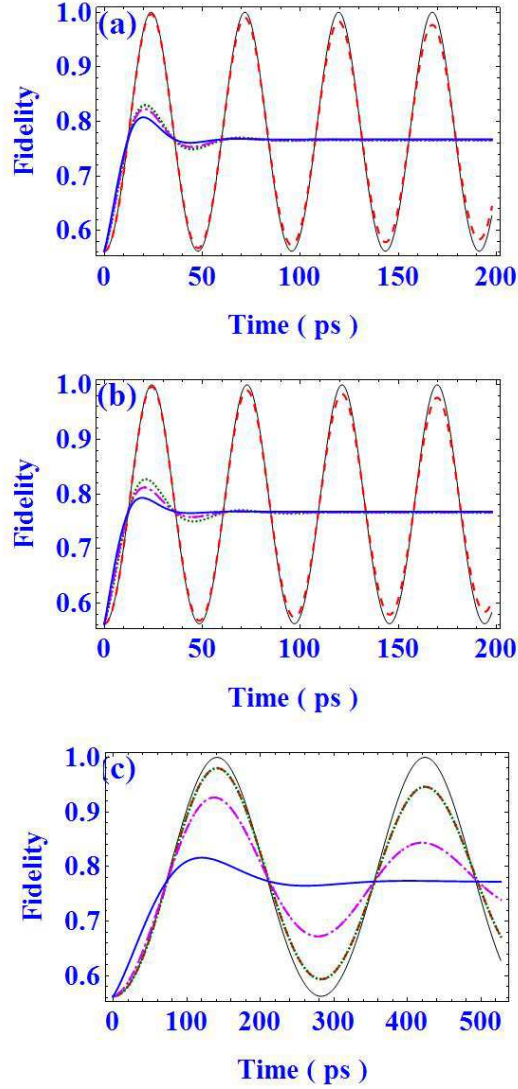}
\caption{(color online). The fidelity of the CPHASE gate in three
cases against delay time(ps) with $\Omega = 0.1$ meV and $2R = 6$ nm
: (a) for ring configuration, (b) for line configuration with
transition $|\Psi_1\rangle\leftrightarrow|\Psi_4\rangle$, (c) for
line configuration with transition
$|\Psi_1\rangle\leftrightarrow|\Psi_2\rangle$. We present five
curves in each figure to describe performance of CPHASE gate
operation in different conditions: without noise(black, solid line),
only spontaneous radiation(red, dashed line), combing both
spontaneous and phonon effect at three finite temperature,
respectively. In (a) and (b),  temperature $T$ varies from $T =
0K$(green, dotted line), $10K$(Pink, dash-dot line), $20K$(blue,
solid line). While, $T$ are chosen to be $0K$(green, dotted
line),$5K$(Pink, dash-dot line),$10K$(blue, solid line) in (c).}
\label{fig:fidelity}
\end{figure}

In the following, we will study two schemes for the phases gate
according to the different configurations of three QDs.

{\it Ring configuration.} --  In the ring-arrayed QDs, we first
reduce the Hamiltonian  Eq.(\ref{eq:Ham}) into the subspace
$\mathcal{H}$. When $\Omega=0$, the eigenstates of $\mathcal{H}$ are
:$|\Psi_1\rangle=|\uparrow\uparrow\uparrow\rangle$,
$|\Psi_2^*\rangle=\frac{1}{\sqrt{6}}(c_{1}^{\dag}-2c_{2}^{\dag}+c_{3}^{\dag})|\uparrow\uparrow\uparrow\rangle$,
$|\Psi_3^*\rangle=-\frac{\sqrt{2}}{2}(c_{1}^{\dag}-c_{3}^{\dag})|\uparrow\uparrow\uparrow\rangle$,
$|\Psi_4\rangle=\frac{1}{\sqrt{3}}(c_{1}^{\dag}+c_{2}^{\dag}+c_{3}^{\dag})|\uparrow\uparrow\uparrow\rangle$.
Considering the case $\Omega\ll V_{F}\ll V_{xx}$, double and ternate
trion states are adiabatically eliminated. Moreover, the
anti-symmetry basis $|\Psi_2^*\rangle$, $|\Psi_3^*\rangle$ have no
effect on the initial state $|\Psi_1\rangle$, so we could only
choose the higher single-exciton state $|\Psi_4\rangle$ as an
auxiliary state $|\Psi_s\rangle$ , and set
$\omega_{l}=\omega_{a}+2V_{F}$. It can be obtained that
$\alpha_{ring} = \sqrt{3}$, and the effective Hamiltonian in
ring-structure is $H_{\text{eff}} = {\sqrt{3}\Omega \over 2}
|\Psi_4\rangle \langle \Psi_1| +h.c $.

In Fig. \ref{fig:fidelity}, we present the fidelity of the CPHASE
gate with $\Omega = 0.1$ meV and $2R =6 $ nm, where the input state
is $|\Psi_{i}\rangle=[\frac{1}{\sqrt{2}}(|\uparrow\rangle+
|\downarrow\rangle)]^{\otimes3}$, and the ideal state is
$|\Psi_{f}\rangle =
\frac{1}{2\sqrt{2}}(|\downarrow\downarrow\downarrow\rangle+|\uparrow\downarrow\downarrow\rangle+|\downarrow\uparrow\downarrow\rangle+|\downarrow\downarrow\uparrow\rangle+|\uparrow\uparrow\downarrow\rangle+|\uparrow\downarrow\uparrow\rangle+|\downarrow\uparrow\uparrow\rangle-|\uparrow\uparrow\uparrow\rangle)$
after the gate operation. The fidelity here is defined as
$\mathcal{F}=|\langle\Psi_{f}|\rho|\Psi_{f}\rangle|$. The black
solid line is the result that set $\Gamma=0$, which shows perfect
Rabi oscillation. Then we consider the phonon effect.  We have
demonstrated that the single-exciton state $|\Psi_2^*\rangle$ and
$|\Psi_3^*\rangle$ are decoupled from the evolution of
$H_{\text{sub}}$ in the case without environment fluctuation.
However, $|\Psi_ 2^*\rangle$ and $|\Psi_3^*\rangle$ are degenerated
with energy lower than $|\Psi_4\rangle$, so spontaneous photon
emission and phonon interaction lead to an intense dissipation and
$|\Psi_ 2^*\rangle$ and $|\Psi_ 3^*\rangle$ should be reconsidered
in our model. The new Hamiltonian is
$H=H_{sub}^{'}+\sum\limits_{\mathbf{q}}\omega_{q}(a_{\mathbf{q}}^{\dag}a_{\mathbf{q}})+H_{ep}$,
where
$H_{sub}^{'}=-V_{F}|\Psi_2^*\rangle\langle\Psi_2^*|-V_{F}|\Psi_3^*\rangle\langle\Psi_3^*|+\frac{\sqrt{3}\Omega}{2}(|\Psi_1\rangle\langle\Psi_4|+h.c.)$.
In the basis consisted by the eigenstates of $H_{sub}^{'}$
($|\Phi_1\rangle=\frac{1}{\sqrt{2}}(|\Psi_1\rangle-|\Psi_4\rangle)$,
$|\Phi_2\rangle=|\Psi_2^*\rangle$,
$|\Phi_3\rangle=|\Psi_3^*\rangle$,
$|\Phi_4\rangle=\frac{1}{\sqrt{2}}(|\Psi_1\rangle+|\Psi_4\rangle)$),
we move the Hamiltonian into an interaction picture with respect to
$H_{sub}^{'}+\sum\limits_{\mathbf{q}}\omega_{q}(a_{\mathbf{q}}a_{\mathbf{q}}^{\dag})$,
and calculate the density operator by approaches mentioned above.
The distance between the center of the ring configuration and each
dot is $d=\sqrt{3}$ nm. We set the coordinates of three dots to be
$(-\frac{d}{2},-\frac{\sqrt{3}d}{2},0)$, $(d,0,0)$ and
$(-\frac{d}{2},\frac{\sqrt{3}d}{2},0)$. The spectral densities
$J(\omega)$ with their relevant parameters shown in Table.
\ref{tab1}.

In order to validate the performance of the proposed phase gate, we
perform a direct numerical simulation with the full master equation
Eq. (\ref{eq:master}). In Fig. \ref{fig:fidelity}, it is shown that
the Rabi oscillations in fidelity are quickly damped primarily due
to phonon-exciton interaction, and no clear oscillation period is
observable even if temperature achieved at its limit $0$K. As
temperature $T$ increases, the system suffers a stronger damping.
The spontaneous radiation is also a decoherent source of damping,
but the influence is so small that Rabi oscillations are obvious in
a relatively long timescale.

\begin{table}[tbp]
\begin{tabular}{c c c c}
\hline $J(\omega)$&  $L$ & $\omega$\\
\hline\hline$8\pi\mathbb{G}(\omega)[1-\frac{\sin(\sqrt{3}|\mathbf{q}|d)}{\sqrt{3}|\mathbf{q}|d}]$&$\frac{1}{2\sqrt{3}}|\Phi_2\rangle\langle\Phi_4|$&$3V_{F}+\frac{1}{2}\alpha\Omega$ \\
&$\frac{1}{2\sqrt{3}}|\Phi_2\rangle\langle\Phi_1|$&$3V_{F}-\frac{1}{2}\alpha\Omega$\\
\hline$3\pi\mathbb{G}(\omega)(|\mathbf{q}|d)^{2}[4-\frac{3}{5}(|\mathbf{q}|d)^{2}]$&$\frac{1}{6}|\Phi_3\rangle\langle\Phi_4|$&$3V_{F}+\frac{1}{2}\alpha\Omega$\\
&$\frac{1}{6}|\Phi_3\rangle\langle\Phi_1|$&$3V_{F}-\frac{1}{2}\alpha\Omega$\\
\hline$3\pi\mathbb{G}(\omega)[12-4(|\mathbf{q}|d)^{2}+\frac{3}{5}(|\mathbf{q}|d)^{4}]$&$\frac{1}{6}|\Phi_1\rangle\langle\Phi_4|$&$\alpha\Omega$\\
\hline
\end{tabular}%
\caption{Calculation results of spectral densities with different
frequencies and dissipative operators in ring configuration. Here
$\mathbb{G}(\omega)= \frac{\omega^{3}}{8\pi^{2}\mu c_{s}^{5}}[D_e
\exp(-(\frac{\omega d_{e}}{2c_{s}})^2)-D_h \exp(-(\frac{\omega
d_{h}}{2c_{s}})^2)]^2$ is the common factor of all spectral
densities in ring structure, $\alpha_{ring} = \sqrt{3}$ is the
character factor in ring configuration and $|\mathbf{q}|=
\omega/c_{s}$.} \label{tab1}
\end{table}

\begin{table}[tbp]
\begin{tabular}{c c c c}
\hline  $J(\omega)$&$L_{down}$&$L_{up}$ & $\omega$
\\ \hline \hline $2\sin(|\mathbf{q}|d)\mathbb{G}(\omega)$&
$\frac{1}{4}|\Phi_1\rangle\langle\Phi_3|$&$\frac{1}{2\sqrt{2}}|\Phi_2\rangle\langle\Phi_3|$&$\sqrt{2}V_{F}+\frac{1}{2}\alpha\Omega$  \\
&$\frac{1}{4}|\Phi_2\rangle\langle\Phi_3|$&$\frac{1}{4}|\Phi_3\rangle\langle\Phi_4|$&$\sqrt{2}V_{F}$ \\
&$\frac{1}{2\sqrt{2}}|\Phi_3\rangle\langle\Phi_4|$&$\frac{1}{4}|\Phi_3\rangle\langle\Phi_1|$&$\sqrt{2}V_{F}-\frac{1}{2}\alpha\Omega$\\
\hline $4\sin^2 (\frac{|\mathbf{q}|d}{2})\mathbb{G}(\omega)$&$\frac{1}{4\sqrt{2}}|\Phi_1\rangle\langle\Phi_4|$&$\frac{1}{4\sqrt{2}}|\Phi_2\rangle\langle\Phi_4|$&$2\sqrt{2}V_{F}+\frac{1}{2}\alpha\Omega$\\
&$\frac{1}{4\sqrt{2}}|\Phi_2\rangle\langle\Phi_4|$&$\frac{1}{4\sqrt{2}}|\Phi_2\rangle\langle\Phi_1|$&$2\sqrt{2}V_{F}-\frac{1}{2}\alpha\Omega$
\\
\hline$4\cos^2(\frac{|\mathbf{q}|d}{2})\mathbb{G}(\omega)$&$\frac{1}{8}|\Phi_1\rangle\langle\Phi_2|$&$\frac{1}{8}|\Phi_1\rangle\langle\Phi_4|$&$\alpha\Omega$
\\
\hline
\end{tabular}%
\caption{Calculation results of spectral densities with different
frequencies and dissipative operators in line configuration. Here
$\mathbb{G}(\omega)= \frac{\omega^{3}}{2\pi\mu c_{s}^{5}}[D_e
\exp(-(\frac{\omega d_{e}}{2c_{s}})^2)-D_h \exp(-(\frac{\omega
d_{h}}{2c_{s}})^2)]^2$ is the common factor of all spectral
densities in line structure, and $|\mathbf{q}|= \omega/c_{s}$. The
character factor $\alpha$ has two different value when choosing
different resonant transitions in the line configuration. If
resonant transition is selected as
$|\Psi_{1}\rangle\leftrightarrow|\Psi_{2}\rangle$, $\alpha$ becomes
as $\alpha_{down} = 1-\frac{\sqrt{2}}{2}$, and relative decay
operator is $L_{down}$; while $\alpha$ becomes as $\alpha_{up} =
1+\frac{\sqrt{2}}{2}$, and decay operator is $L_{up}$ if
$|\Psi_{1}\rangle\leftrightarrow|\Psi_{4}\rangle$ is selected. }
\label{tab2}
\end{table}

{\it Line Configuration.}-- The analysis of the line configuration
can be derived in analog to the ring-configuration. In
subspace $\mathcal{H}$, eigenstates for $\Omega=0$ include
$|\Psi_1\rangle=|\uparrow\uparrow\uparrow\rangle$,
$|\Psi_2\rangle=\frac{1}{2}(c_{1}^{\dagger}-\sqrt{2}c_{2}^{\dagger}+
c_{3}^{\dagger})|\uparrow\uparrow\uparrow\rangle$,
$|\Psi_3^\ast\rangle=
-\frac{\sqrt{2}}{2}(c_{1}^{\dagger}-c_{3}^{\dagger})|\uparrow\uparrow\uparrow\rangle$,
and
$|\Psi_4\rangle=\frac{1}{2}(c_{1}^{\dagger}+\sqrt{2}c_{2}^{\dagger}+
c_{3}^{\dagger})|\uparrow\uparrow\uparrow\rangle$. The energy level
diagram is illustrated in Fig.\ref{fig:scheme}(d). It is shown that
 both $|\Psi_2\rangle$ and
$|\Psi_4\rangle$ have an energy shift comparing with other single
exciton state, and can be chosen as auxiliary level resonated with
$|\Psi_1\rangle$ to prepare a CPHASE Gate.

 If we pump the laser with
frequency $\omega_{l}=\omega_{a}-\sqrt{2}V_{F}$, the transition
$|\Psi_{1}\rangle\leftrightarrow|\Psi_{2}\rangle$ is allowed (in
Fig. \ref{fig:scheme}). We can obtain the character factor of
configuration $\alpha_{down} = 1-\frac{\sqrt{2}}{2}$,
$H_{eff}(t)=\frac{\Omega}{2}(1-\frac{\sqrt{2}}{2})(|\Psi_
1\rangle\langle\Psi_ 2|+H.c.)$. In the case of
$\int_{0}^{t_{I}}\frac{\Omega}{2}(1-\frac{\sqrt{2}}{2}) dt=\pi$, a
$\pi$-pulse shift on the system engenders:
$|\uparrow\uparrow\uparrow\rangle\rightarrow-|\uparrow\uparrow\uparrow\rangle$.

If we let the laser be resonant with transition
$|\Psi_{1}\rangle\leftrightarrow|\Psi_{4}\rangle$ (simply replace
$\omega_{l}$ by $\omega_{l}=\omega_{a}+\sqrt{2}V_{F}$ ), the factor
$\alpha$ becomes as $\alpha_{up} = 1+\frac{\sqrt{2}}{2}$, a similar
Hamiltonian is
$H_{eff}(t)=\frac{\Omega}{2}(1+\frac{\sqrt{2}}{2})(|\Psi_
1\rangle\langle\Psi_ 2|+H.c.)$, and the $\pi$-pulse shift may occur
when $\int_{0}^{t_{I}}\frac{\Omega}{2}(1+\frac{\sqrt{2}}{2}) dt=\pi$
and the operation time could be obviously shortened.

Like the ring configuration, we can also derive a new density
operator to describe the evolution of line configuration. The
distance between the center of the ring configuration and each dot
is $2d=6$ nm. Here the coordinates of three dots are chosen as $(-d,
0, 0)$, $(0, 0, 0)$ and $(d, 0, 0)$, then we can obtain spectral
densities $J(\omega)$(listed in Table. \ref{tab2}) in two strategies
about the selection of resonant transition from ground state to
$|\Psi_2\rangle$ (\textit{low-level}) or high exciton energy level
$|\Psi_4\rangle$(\textit{high-level}) , and then we may find that
two numerical calculation results in the same configuration is
totally different. In Fig. \ref{fig:fidelity}, it is demonstrated
that, in the \textit{low-level} case, phonon effect does not lead a
strong damping at very low temperature. Especially at $0$K, the
behavior is as the same as that only considering spontaneous
radiation. This is because spectral density $J(\omega)$ between
ground state and low single exciton level is close to zero. As the
temperature is increased, oscillations are damped: for $T = 5$K, two
periods of oscillation are still visible, while no clear oscillation
is exhibited for $T = 10$K. In the \textit{high-level} case, like
the ring configuration, phonon effect plays a dominant role in
decoherence mechanism. Moreover, the damping induced by phonon
effect is a little larger than ring configuration at the same
temperature.

{\it Conclusion.}--In this work, we discuss the configuration effect
and acoustic phonon effect on the CPHASE gate. We found that the
quality of the phase gate depends strongly on the configuration,
which is of great importance when QDs is used as building block to
larger integrated systems. We also found that those configuration
mainly influence the energy levels, hence influence the exciton
phonon coupling strength. We show that the line configuration with
resonant transition $|\Psi_1\rangle\leftrightarrow|\Psi_2\rangle$ is
more suitable for quantum computation.

S. Y would like to thank Dr. Erik. M. Gauger for his kind and warm
help by Email. This work was supported by National Fundamental
Research Program, also by National Natural Science Foundation of
China (Grant No. 10674128 and 60121503) and the Innovation Funds and
\textquotedblleft Hundreds of Talents\textquotedblright\ program of
Chinese Academy of Sciences and Doctor Foundation of Education
Ministry of China (Grant No. 20060358043).

\end{document}